# APPLICATION OF POLYNOMIAL VECTOR (PV) PROCESSING TO IMPROVE THE ESTIMATION PERFORMANCE OF BIO DIESEL IN VARIABLE COMPRESSION RATIO DIESEL ENGINE

Suresh M.,
Asst.Prof,Mechanical Engineering, Sri Sai Ram Engg. College,Chennai-44,Tamilnadu,India

Maheswar Dutta
Professor and Principal, M.N.R Engg. College, Hyderabad, India

Purushothaman S
Professor and Dean, Mechanical Engineering, Udaya School of Engineering, India-629204
drsppuru@gmail.com

*Abstract-*This paper presents the implementation of polynomial vector back propagation algorithm (PVBPA) for estimating the power, torque, specific fuel consumption and presence of carbon monoxide, hydrocarbons in the emission of a direct injection diesel engine. Experimental readings were obtained using the biodiesel prepared form the waste low quality cooking oil collected from the canteen of Sri Sairam Engineering College, India.. This waste cooking oil was due to the preparation of varieties of food (vegetables fried and non vegetarian). Over more than a week, trans esterification was done in chemical lab and the biodiesel was obtained. The biodiesel was mixed in proportions of 10%, 20 % , 30%,40%, 50% with remaining combinations of the diesel supplied by the Indian government. Variable compression ratio (VCR) diesel engine with single cylinder, four stroke diesel type was used. The outputs of the engine as power, torque and specific fuel consumption were obtained from the computational facility attached to the engine. The data collected for different input conditions of the engine was further used to train (PVBPA). The trained PVBPA network was further used to predict the power, torque and brake specific fuel consumption (SFC) for different speed, biodiesel and diesel combinations and full load condition. The estimation performance of the PVBPA network is discussed.

*Keywords: polynomial vector, back propagation algorithm, waste cooking oil, biodiesel.*

I   INTRODUCTION

In this paper, performance of a diesel engine and exhaust emission content of the diesel engine when using Biodiesel blended with diesel has been analyzed. Data collected from the engine for various loads / speed were used to train polynomial vector back propagation (PVBPA) neural networks.. Subsequently, the PVBPA was used to estimate the performance of the diesel engine and estimate the quality of the exhaust gas for different loads / speeds and combinations of fuel other than that used for training of the PVBPA.

Biodiesel refers to a vegetable oil or animal based diesel fuel consisting of long chain (methyl, propyl or ethyl) esters. Biodiesel [1-3] is typically made by chemically reacting lipids (eg. Vegetable oil, animal fat, tallow) with an alcohol producing fatty acid esters. The various Multipurpose oils [8,10-12]also used as biofuel such as Castor oil, Coconut oil (copra oil), Colza oil, Corn oil, Cottonseed oil, False flax oil, Hemp oil, Mustard oil, Palm oil, Peanut oil, Radish oil. Rapeseed oil, Ramtil oil, Rice bran oil, Safflower oil, Salicornia oil, Soybean oil, Sunflower oil, Tigernut oil , Tung oil, are lists of vegetable oils that are suitable for biodiesel. Similarly, Inedible oils used only or primarily as biofuel such as Copaiba, Honge oil, Jatropha oil, Jojoba oil, Milk bush, Nahor oil, Paradise oil, Petroleum nut oil.

Vegetable oils are evaluated for use as a biofuel based on: a) Suitability as a fuel, based on flash point, energy content, viscosity, combustion products and other factors, b) Cost, based in part on yield, effort required to grow and harvest, and post-harvest processing cost.

Alternative fuels for diesel engines are becoming increasingly important due to diminishing petroleum reserves and the environmental consequences of exhaust gases from petroleum fuelled engines. A number of studies have shown that triglycerides hold promise as alternative diesel engine fuels. So, many countries are interested in that.

## II EXPERIMENTAL INVESTIGATION

The setup consists of single cylinder, four stroke, VCR (Variable Compression Ratio) Diesel engine connected to eddy current type dynamometer for loading. The compression ratio can be changed without stopping the engine and without altering the combustion chamber geometry by specially designed tilting cylinder block arrangement. Setup is provided with necessary instruments for combustion pressure measurements. The setup has stand-alone panel box consisting of air box, two fuel tanks for duel fuel test, manometer, fuel measuring unit, transmitters for air and fuel flow measurements, process indicator and engine indicator. Rotameters are provided for cooling water and calorimeter water flow measurement.

The setup enables study of VCR engine performance for brake power, indicated power, frictional power, brake mean effective pressure (BMEP), indicated mean effective pressure (IMEP), brake thermal efficiency, indicated thermal efficiency, Mechanical efficiency, volumetric efficiency, specific fuel consumption, A/F ratio and heat balance. Labview based Engine Performance Analysis software package "EnginesoftLV" is provided for on line performance evaluation.

1. Brake power (BP)= $2 * \pi * n \; T / (60 * 1000)$
2. Brake specific fuel consumption (Kg/kwh)= Fuel flow in kg / hour / BP
3. Specific fuel consumption (SFC): Brake specific fuel consumption and indicated specific fuel consumption, abbreviated BSFC and ISFC, are the fuel consumptions on the basis of Brake power and Indicated power respectively.

### 2.1. Biodiesel preparation

In the present investigation, biodiesel was produced from waste cooking oil from the canteen of

Sri Sairam Engineering College, India. 2 gram Alkali catalyst and 35 cc methanol (as an alcohol) was applied for 150 gram waste cooking oil in this reaction. Biodiesel production reaction time was two hour with stirring and with moderate heat. Upto two weeks time is needed for separation. The waste cooking oil methyl ester was added to diesel fuel in 10 to 50 percent ratios and then used as fuel for one cylinder diesel engine.

*2.2. Experimental set up and test procedure*

The experimental setup consists of single cylinder diesel engine, an engine test bed and a gas analyzer. The engine setup is shown in Figure 1. The schematic of the experimental setup is shown in Figure 2.

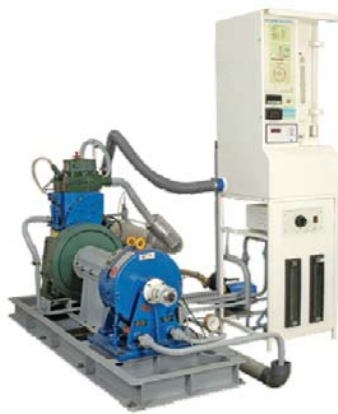

Figure 1. Variable compression ratio diesel engine (Apex innovations)

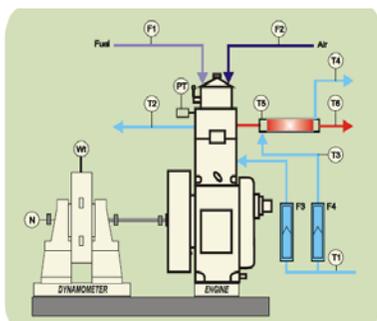

Figure 2. Schematic layout of the setup (Apex innovations)

where

F1 Fuel consumption kg/hr
F2 Air consumption kg/hr
F3,F4 Calorimeter water flow kg/hr
T3 Calorimeter water inlet temperature $^{o}K$
T2,T4 Calorimeter water outlet temperature $^{o}K$
T5 Exhaust gas to calorimeter inlet temp. $^{o}K$
T6 Exhaust gas from calorimeter outlet temp. $^{o}K$

There are two fuel tanks, one is for diesel fuel and the other for fuel blends. The engine under study is a VCR, water cooled single cylinder, in-line, naturally aspirated, Kirlosker diesel engine. The test engine was coupled to an electric eddy current dynamometer. A vehicle gas analyzer model was used for measuring CO and HC emissions. Engine was run at several speeds at full load and power, torque, fuel consumption and emissions was measured. Table 1 presents experimental data obtained. Table 2 presents the CO and HC emissions value obtained from exhaust gas analyzer.

| TABLE I. | | | | | | | | |
|---|---|---|---|---|---|---|---|---|
| S.No. | Full load | Waste cooking oil (Biodiesel) | Diesel | Speed(rpm) of the engine | Power (KW) | Torque (N-M) | Specific fuel consumption (SFC) Litre/KW |
| 1 | 1 | 0 | 1 | 1200 | 6.2 | 48 | 0.32 |
| 2 | 1 | 0 | 1 | 1600 | 9.2 | 54 | 0.33 |
| 3 | 1 | 0 | 1 | 2000 | 12.3 | 57 | 0.34 |
| 4 | 1 | 0 | 1 | 2400 | 16.0 | 63 | 0.33 |
| 5 | 1 | 0 | 1 | 2800 | 17.6 | 53 | 0.33 |
| 6 | 1 | 0 | 1 | 3200 | 17.7 | 51 | 0.34 |
| 7 | 1 | 10 | 90 | 1200 | 7.0 | 54 | 0.34 |
| 8 | 1 | 10 | 90 | 1600 | 9.8 | 57 | 0.35 |
| 9 | 1 | 10 | 90 | 2000 | 12.0 | 56 | 0.33 |
| 10 | 1 | 10 | 90 | 2400 | 15.2 | 62 | 0.30 |
| 11 | 1 | 10 | 90 | 2800 | 16.1 | 55 | 0.31 |
| 12 | 1 | 10 | 90 | 3200 | 16.3 | 48 | 0.37 |
| 13 | 1 | 20 | 80 | 1200 | 6.6 | 51 | 0.33 |
| 14 | 1 | 20 | 80 | 1600 | 9.2 | 53 | 0.33 |
| 15 | 1 | 20 | 80 | 2000 | 12.8 | 55 | 0.30 |
| 16 | 1 | 20 | 80 | 2400 | 16.3 | 58 | 0.29 |

| 17 | 1 | 20 | 80 | 2800 | 16.8 | 54 | 0.32 |
| 18 | 1 | 20 | 80 | 3200 | 18.0 | 52 | 0.33 |
| 19 | 1 | 30 | 70 | 1200 | 6.8 | 47 | 0.32 |
| 20 | 1 | 30 | 70 | 1600 | 9.6 | 51 | 0.31 |
| 21 | 1 | 30 | 70 | 2000 | 12.4 | 57 | 0.29 |
| 22 | 1 | 30 | 70 | 2400 | 15.0 | 64 | 0.34 |
| 23 | 1 | 30 | 70 | 2800 | 16.8 | 59 | 0.33 |
| 24 | 1 | 30 | 70 | 3200 | 17.4 | 48 | 0.36 |
| 25 | 1 | 40 | 60 | 1200 | 6.0 | 52 | 0.32 |
| 26 | 1 | 40 | 60 | 1600 | 9.6 | 56 | 0.31 |
| 27 | 1 | 40 | 60 | 2000 | 12.4 | 58 | 0.27 |
| 28 | 1 | 40 | 60 | 2400 | 15.0 | 59 | 0.31 |
| 29 | 1 | 40 | 60 | 2800 | 18.0 | 56 | 0.32 |
| 30 | 1 | 40 | 60 | 3200 | 17.0 | 53 | 0.34 |
| 31 | 1 | 50 | 50 | 1200 | 6.2 | 48 | 0.31 |
| 32 | 1 | 50 | 50 | 1600 | 9.0 | 53 | 0.32 |
| 33 | 1 | 50 | 50 | 2000 | 12.4 | 56 | 0.33 |
| 34 | 1 | 50 | 50 | 2400 | 15.8 | 59 | 0.32 |
| 35 | 1 | 50 | 50 | 2800 | 17.0 | 58 | 0.33 |
| 36 | 1 | 50 | 50 | 3200 | 16.8 | 50 | 0.36 |

| TABLE II. | Exhaust gas analyzer output | |
|---|---|---|
| **Fuel blend** | **HC** | **C** |
| B0 | 32 | 0.48 |
| B10 | 18 | 0.49 |
| B20 | 16 | 0.46 |
| B30 | 5 | 0.45 |
| B40 | 7 | 0.4 |
| B50 | 7 | 0.38 |

### III POLYNOMIAL INTERPOLATION

The experimental data presented in Table 1 are further processed to make the data orthogonal to each other. The input vector is pre-processed and then presented to the network. The pre-processing generates a polynomial decision boundary. The pre-processing of the input vector is done as follows:

Let X present the normalized input vector, where

$X = \{X_i\}$; i=1,…nf,

$X_i$ is the feature of the input vector, and nf is the number of features (nf = 4)

An outer product matrix $X_{op}$ of the original input vector is formed, and it is given by:

$$X_{op.} = \begin{bmatrix} X1X1 & X1X2 & X1X3 & X1X4 \\ X2X1 & X2X2 & X2X3 & X2X4 \\ X3X1 & X3X2 & X3X3 & X3X4 \\ X4X1 & X4X2 & X4X3 & X4X4 \end{bmatrix}$$

Using the $X_{op}$ matrix, the following polynomials are generated:

1) Product of inputs (NL1)

it is denoted by:

$\Sigma w_{ij}x_i$ (i≠j) = Off-diagonal elements of the outer product matrix.

The pre-processed input vector is a 6-dimensional vector.

2) Quadratic terms (NL2)

It is denoted by:

$\Sigma w_{ij}x_{i2}$ = Diagonal elements of the outer product matrix.

The pre-processed input vector is a 4-dimensional vector.

3) A combination of product of inputs and quadratic terms (NL3)

It is denoted by:

$\Sigma w_{ij}x_i(i≠j) + \Sigma w_{ij}x_{i2}$ = Diagonal elements and Off-diagonal elements of the outer product matrix. The pre-processed input vector is a 10(6+4) dimensional vector.

4) Linear plus NL1 (NL4)

The pre-processed input vector is a 10-dimensional vector.

5) Linear plus NL2 (NL5)

The pre-processed input vector is a 8-dimensional vector.

6) Linear plus NL3 (NL6)

The pre-processed input vector is a 14-dimensional vector.

In the above polynomials such as NL4, NL5 and NL6

vector, the term 'linear' represents the normalized input pattern without pre-processing.

When the training of the network is done with a fixed pre-processing of the input vector, the number of iterations required is less than that required for the training of the network without pre-processing of the input vector to reach the desired MSE.

The combinations of different pre-processed methods with different synaptic weight update algorithms are shown in Table 3. As shown in Table 3, BPA weight update algorithms have been used with fixed pre-processed input vectors for learning.

| TABLE III Combination of BPA with different pre-processed input vectors | |
|---|---|
| BPA + NL1 | BPA + NL2 |
| BPA + NL3 | BPA + NL4 |
| BPA + NL5 | BPA + NL6 |
| NL is non-linear , 1-6 are the types | |

## IV Back propagation Algorithm

A neural network is constructed by highly interconnected processing units (nodes or neurons) which perform simple mathematical operations, [5]. Neural networks are characterized by their topologies, weight vectors and activation function which are used in the hidden layers and output layer, [9]. The topology refers to the number of hidden layers and connection between nodes in the hidden layers. The activation functions that can be used are sigmoid, hyperbolic tangent and sine. The network models can be static or dynamic [7]. Static networks include single layer perceptrons and multilayer perceptrons. A perceptron or adaptive linear element (ADALINE), [4,6], refers to a computing unit. This forms the basic building block for neural networks. The input to a perceptron is the summation of input pattern vectors by weight vectors. In most of the applications one hidden layer is sufficient. The activation function which is used to train the ANN, is the sigmoid function.

TRAINING STEPS INVOLVED.

*Forward propagation*

**Step 1:** The weights of the network are initialized.

**Step 2:** The inputs and outputs of a pattern are presented to the network.

**Step 3:** The output of each node in the successive layers is calculated.

$$o_{(output\ of\ a\ node)} = 1/(1+\exp(-\sum w_{ij} x_i)) \quad (1)$$

**Step 4:** The error of a pattern is calculated

$$E(p) = (1/2) \sum (d(p) - o(p))^2 \quad (2)$$

*Reverse propagation*

**Step 1:** The error for the nodes in the output layer is calculated.

$$\delta_{(output\ layer)} = o(1-o)(d-o) \quad (3)$$

**Step 2:** The weights between output layer and hidden layer are updated.

$$W(n+1) = W(n) + \eta \delta_{(output\ layer)} o_{(hidden\ layer)} \quad (4)$$

**Step 3:** The error for the nodes in the hidden layer is calculated

$$\delta_{(hidden\ layer)} = o(1-o) \sum \delta_{(output\ layer)} W_{(updated\ weights\ between\ hidden\ and\ output\ layer)} \quad (5)$$

**Step 4:** The weights between hidden and input layer are updated.

$$W(n+1) = W(n) + \eta \delta_{(hidden\ layer)} o_{(input\ layer)} \quad (6)$$

Where

o is the actual output of a node in hidden or output layer.

$\eta$ is the learning factor.

$\delta$ is the error of node.

P is the pattern number.

E is the errors of nodes in the output layer for a pattern.

The above steps complete one weight updation. Second pattern is presented and the above steps are followed for the second weight updation. When all the training patterns are presented, a cycle of iteration or epoch is completed. The errors of all the training patterns are calculated and displayed on the monitor as the mean squared error (MSE).

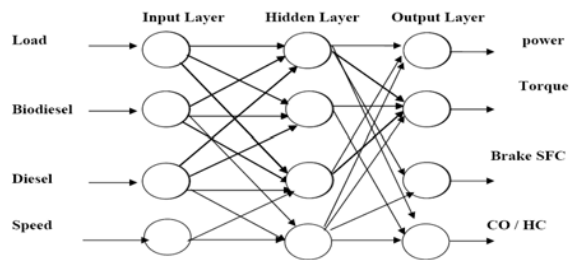

Figure 3. PVBPA network for predicting engine performance

## V. RESULTS AND DISCUSSION

*4.1 Properties of waste cooking oil biodiesel fuels*

| Property | Biodiesel |
|---|---|
| Flash point, closed cup | 182 °C |
| Pour point | -3°C |
| Kinematical viscosity, 40°C | 4.15 mm$^2$/s |
| Total Sulfur | 0.0018 wt. % |
| Copper strip corrosion | 1a |
| Cloud point | 0 °C |

*4.2. Torque and Power*

Fuel rack is placed in maximum fuel injection position for full load conditions. The engine is loaded slowly. The engine speed is reduced with increasing load. Range of speed was selected between 1200 – 3600 rpm. Engine test results with net diesel fuel showed that maximum torque was 64.2 Nm which occurred at 2400 rpm. The maximum power was 18.12 kW at 3200 rpm. Power and torque for fuel blends at full load is shown in Table 1. The power estimation by different BPA with NL combinations are shown in Figures 4-9. The torque estimation by BPA with different NL are presented in Figures 10-15.

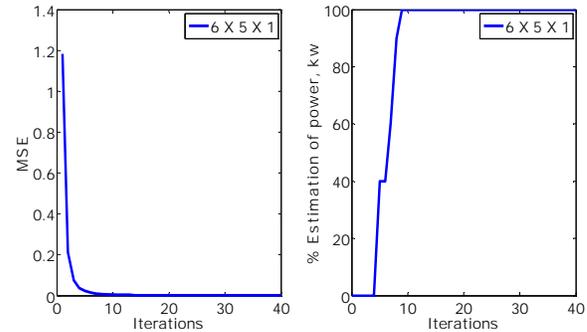

Figure 4. Estimation of power by BPA+ NL1

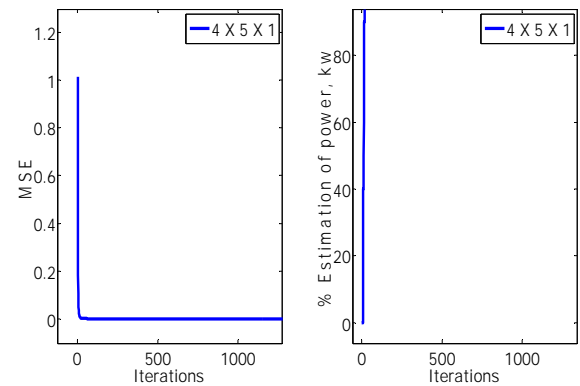

Figure 5. Estimation of power by BPA+ NL2

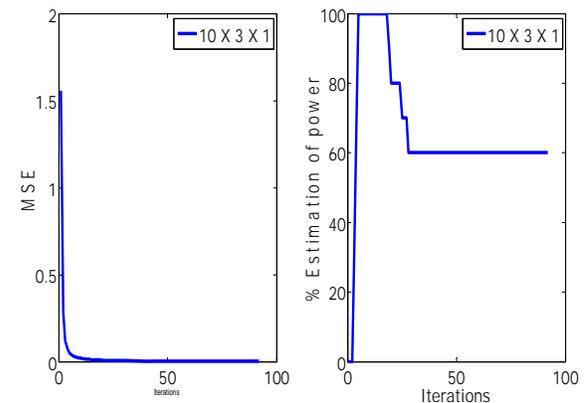

Figure 6. Estimation of power by BPA+ NL3

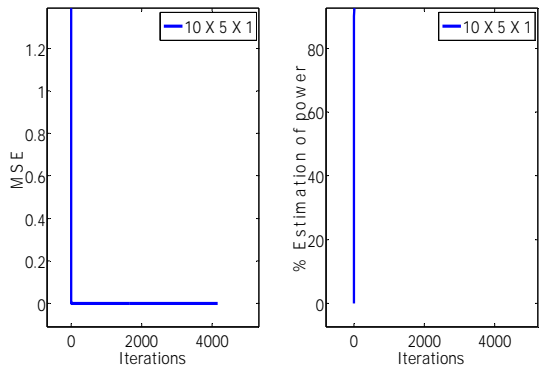

Figure 7.    Estimation of power by BPA+ NL4

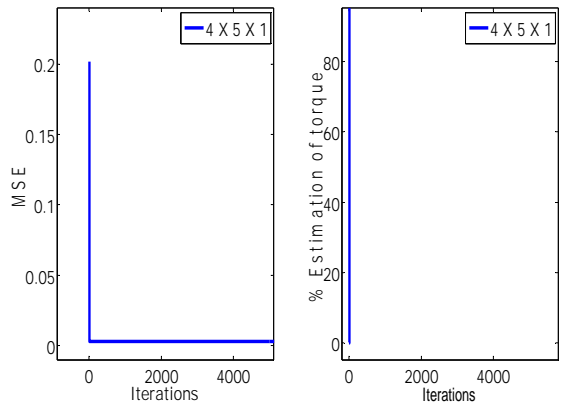

Figure 11.    Estimation of torque by BPA+NL2

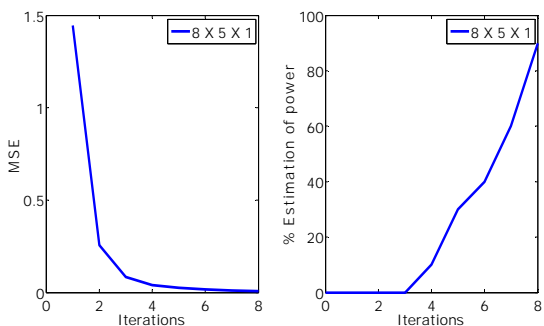

Figure 8.    Estimation of power by BPA+ NL5

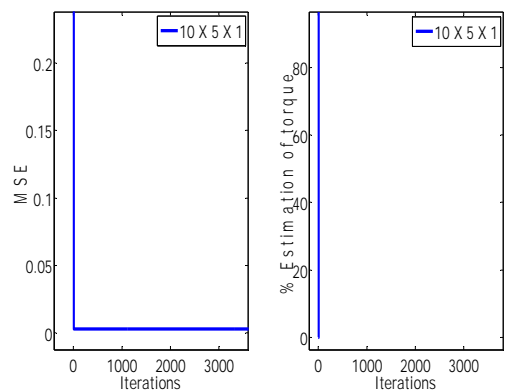

Figure 12.    Estimation of torque by BPA+NL2

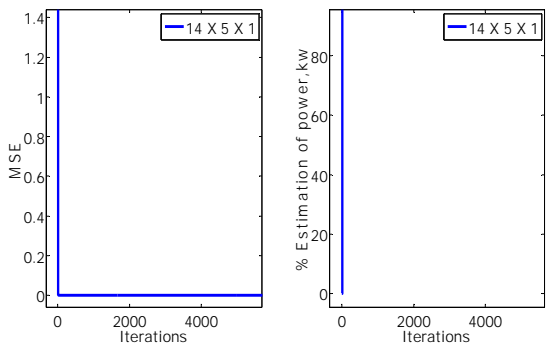

Figure 9.    Estimation of power by BPA+ NL6

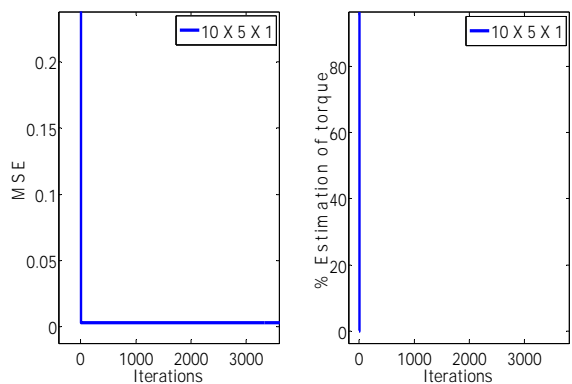

Figure 13.    Estimation of torque by BPA+NL4

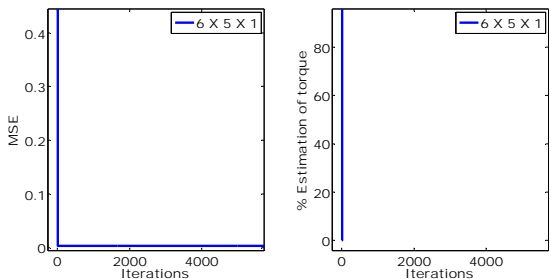

Figure 10.    Estimation of torque by BPA+NL1

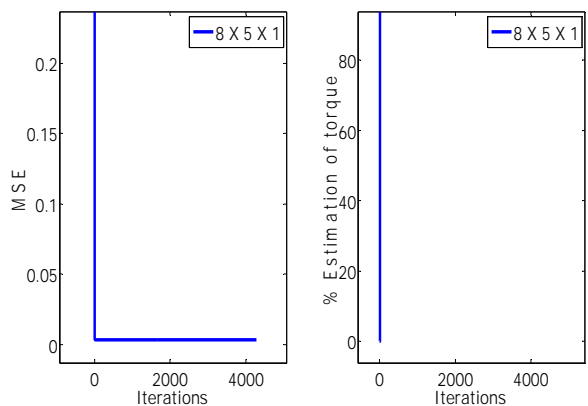

Figure 14.  Estimation of torque by BPA+NL5

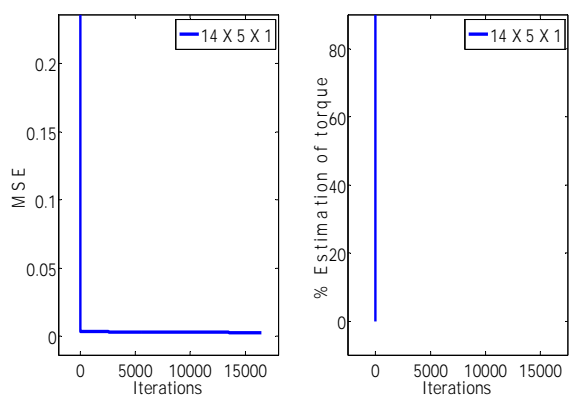

Figure 15.  Estimation of torque by BPA+NL6

## VI. CONCLUSION

The engine has been tested under same operating conditions with diesel fuel and waste cooking biodiesel fuel blends. The results were found to be very comparable. The maximum power and torque produced using diesel fuel was 18.2 kW and 64.2 Nm at 3200 and 2400 rpm respectively. By adding 20% of waste cooking oil methyl ester, the maximum power and torque increased by 2.7% and 2.9% respectively. The concentration of the CO and HC emissions were significantly decreased when biodiesel was used (Table 2).